 \documentclass{vgtc}                          

\usepackage[table,xcdraw,svgnames,dvipsnames]{xcolor}
\usepackage{tikz}
\usepackage{enumitem}

\graphicspath{{figures/}{pictures/}{images/}{./}} 

\usepackage{times}                     

\usepackage{tabu}                      
\usepackage{booktabs}                  
\usepackage{lipsum}                    
\usepackage{mwe}                       

\usepackage{stfloats}

\usepackage{mathptmx}                  

\newcommand\I[1]{\textit{#1}}

\newcommand\eat[1]{}

\usepackage{multirow}

\usepackage{collcell}
\usepackage{xstring}
\definecolor{tabred}{RGB}{204,0,0}
\definecolor{tabyellow}{RGB}{225,229,153}
\definecolor{tabgreen}{RGB}{106,168,79}
\newcommand*{\WhiteTextCutoffNumberHalf}{59}
\newcommand*{\WhiteTextCutoffNumberFull}{20}
\newcommand*{\MinHalfNumber}{50}
\newcommand*{\MidHalfNumber}{75}
\newcommand*{\MaxHalfNumber}{100}
\newcommand*{\MinFullNumber}{0}
\newcommand*{\MidFullNumber}{50}
\newcommand*{\MaxFullNumber}{100}

\newcommand{\ApplyHalfGradient}[1]{%
    \exploregroups
    \expandarg
    \StrFindGroup{#1}{1}[\ingroup]
    \StrRemoveBraces{\ingroup}[\clean]
    
    \IfDecimal{#1}{
            \ifdim \integerpart pt > \MidHalfNumber pt
                \ifdim \integerpart pt > \WhiteTextCutoffNumberHalf pt
                    \pgfmathsetmacro{\PercentColor}{max(min(100.0*(\integerpart - \MidHalfNumber)/(\MaxHalfNumber-\MidHalfNumber),100.0),0.00)} %
                    \edef\x{\noexpand\cellcolor{tabgreen!\PercentColor!tabyellow}}\x{#1}
                \else
                    \pgfmathsetmacro{\PercentColor}{max(min(100.0*(\integerpart - \MidHalfNumber)/(\MaxHalfNumber-\MidHalfNumber),100.0),0.00)} %
                    \edef\x{\noexpand\cellcolor{tabgreen!\PercentColor!tabyellow}}\x{\textcolor{white}{#1}}
                \fi
            \else
                \ifdim \integerpart pt > \WhiteTextCutoffNumberHalf pt
                    \pgfmathsetmacro{\PercentColor}{max(min(100.0*(\MidHalfNumber - \integerpart)/(\MidHalfNumber-\MinHalfNumber),100.0),0.00)} %
                    \edef\x{\noexpand\cellcolor{tabred!\PercentColor!tabyellow}}\x{#1}   
                \else
                    \pgfmathsetmacro{\PercentColor}{max(min(100.0*(\MidHalfNumber - \integerpart)/(\MidHalfNumber-\MinHalfNumber),100.0),0.00)} %
                    \edef\x{\noexpand\cellcolor{tabred!\PercentColor!tabyellow}}\x{\textcolor{white}{#1}}   
                \fi
            \fi
    }
    {\IfDecimal{\clean}{
            \ifdim \integerpart pt > \MidHalfNumber pt
                \ifdim \integerpart pt > \WhiteTextCutoffNumberHalf pt
                    \pgfmathsetmacro{\PercentColor}{max(min(100.0*(\integerpart - \MidHalfNumber)/(\MaxHalfNumber-\MidHalfNumber),100.0),0.00)} %
                    \edef\x{\noexpand\cellcolor{tabgreen!\PercentColor!tabyellow}}\x{#1}
                \else
                    \pgfmathsetmacro{\PercentColor}{max(min(100.0*(\integerpart - \MidHalfNumber)/(\MaxHalfNumber-\MidHalfNumber),100.0),0.00)} %
                    \edef\x{\noexpand\cellcolor{tabgreen!\PercentColor!tabyellow}}\x{\textcolor{white}{#1}}
                \fi
            \else
                \ifdim \integerpart pt > \WhiteTextCutoffNumberHalf pt
                    \pgfmathsetmacro{\PercentColor}{max(min(100.0*(\MidHalfNumber - \integerpart)/(\MidHalfNumber-\MinHalfNumber),100.0),0.00)} %
                    \edef\x{\noexpand\cellcolor{tabred!\PercentColor!tabyellow}}\x{#1}
                \else
                    \pgfmathsetmacro{\PercentColor}{max(min(100.0*(\MidHalfNumber - \integerpart)/(\MidHalfNumber-\MinHalfNumber),100.0),0.00)} %
                    \edef\x{\noexpand\cellcolor{tabred!\PercentColor!tabyellow}}\x{\textcolor{white}{#1}}
                \fi
            \fi
    }{#1}}
}

\newcommand{\ApplyFullGradient}[1]{%
    \exploregroups
    \expandarg
    \StrFindGroup{#1}{1}[\ingroup]
    \StrRemoveBraces{\ingroup}[\clean]
    
    \IfDecimal{#1}{
            \ifdim \integerpart pt > \MidFullNumber pt
                \ifdim \integerpart pt > \WhiteTextCutoffNumberFull pt
                    \pgfmathsetmacro{\PercentColor}{max(min(100.0*(\integerpart - \MidFullNumber)/(\MaxFullNumber-\MidFullNumber),100.0),0.00)} %
                    \edef\x{\noexpand\cellcolor{tabgreen!\PercentColor!tabyellow}}\x{#1}
                \else
                    \pgfmathsetmacro{\PercentColor}{max(min(100.0*(\integerpart - \MidFullNumber)/(\MaxFullNumber-\MidFullNumber),100.0),0.00)} %
                    \edef\x{\noexpand\cellcolor{tabgreen!\PercentColor!tabyellow}}\x{\textcolor{white}{#1}}
                \fi
            \else
                \ifdim \integerpart pt > \WhiteTextCutoffNumberFull pt
                    \pgfmathsetmacro{\PercentColor}{max(min(100.0*(\MidFullNumber - \integerpart)/(\MidFullNumber-\MinFullNumber),100.0),0.00)} %
                    \edef\x{\noexpand\cellcolor{tabred!\PercentColor!tabyellow}}\x{#1}
                \else
                    \pgfmathsetmacro{\PercentColor}{max(min(100.0*(\MidFullNumber - \integerpart)/(\MidFullNumber-\MinFullNumber),100.0),0.00)} %
                    \edef\x{\noexpand\cellcolor{tabred!\PercentColor!tabyellow}}\x{\textcolor{white}{#1}}
                \fi
            \fi
    }
    {\IfDecimal{\clean}{
            \ifdim \integerpart pt > \MidFullNumber pt
                \ifdim \integerpart pt > \WhiteTextCutoffNumberFull pt
                    \pgfmathsetmacro{\PercentColor}{max(min(100.0*(\integerpart - \MidFullNumber)/(\MaxFullNumber-\MidFullNumber),100.0),0.00)} %
                    \edef\x{\noexpand\cellcolor{tabgreen!\PercentColor!tabyellow}}\x{#1}
                \else
                    \pgfmathsetmacro{\PercentColor}{max(min(100.0*(\integerpart - \MidFullNumber)/(\MaxFullNumber-\MidFullNumber),100.0),0.00)} %
                    \edef\x{\noexpand\cellcolor{tabgreen!\PercentColor!tabyellow}}\x{\textcolor{white}{#1}}
                \fi
            \else
                \ifdim \integerpart pt > \WhiteTextCutoffNumberFull pt
                    \pgfmathsetmacro{\PercentColor}{max(min(100.0*(\MidFullNumber - \integerpart)/(\MidFullNumber-\MinFullNumber),100.0),0.00)} %
                    \edef\x{\noexpand\cellcolor{tabred!\PercentColor!tabyellow}}\x{#1}
                \else
                    \pgfmathsetmacro{\PercentColor}{max(min(100.0*(\integerpart - \MidFullNumber)/(\MaxFullNumber-\MidFullNumber),100.0),0.00)} %
                    \edef\x{\noexpand\cellcolor{tabgreen!\PercentColor!tabyellow}}\x{\textcolor{white}{#1}}
                \fi
            \fi
    }{#1}}
}

\newcolumntype{R}{>{\collectcell\ApplyHalfGradient}c<{\endcollectcell}}

\newcolumntype{S}{>{\small\collectcell\ApplyHalfGradient}c<{\endcollectcell}}

\newcolumntype{T}{>{\small\collectcell\ApplyHalfGradient}c<{\endcollectcell}}

\onlineid{8519}

\vgtccategory{Research}

\vgtcinsertpkg

\title{Virtual Reality Games: Extending Unity Learn Games to VR}

\author{Ryan P. McMahan\thanks{e-mail: rpm@vt.edu}\\ %
    \scriptsize Virginia Tech
\and Nayan N. Chawla\thanks{e-mail: nnchawla@vt.edu}\\ %
    \scriptsize Virginia Tech
\and Christian S. Cassell\thanks{e-mail: ccassell24@vt.edu}\\ %
    \scriptsize Virginia Tech
\and Christopher Peerapon Lee\thanks{e-mail: chrislee24@vt.edu}\\ %
    \scriptsize Virginia Tech}

\teaser{
  \centering
  \includegraphics[width=.49\linewidth]{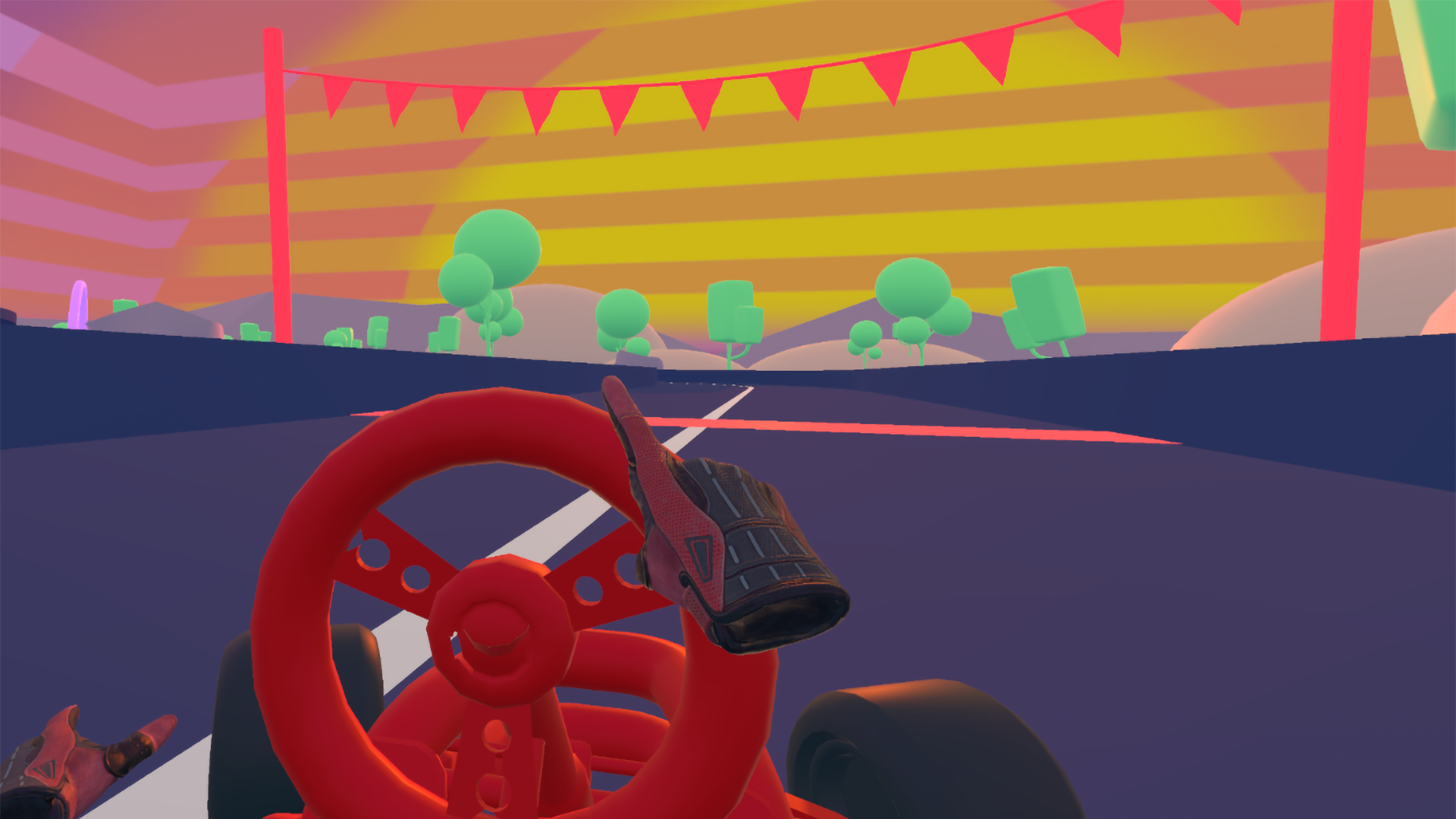}
  \includegraphics[width=.49\linewidth]{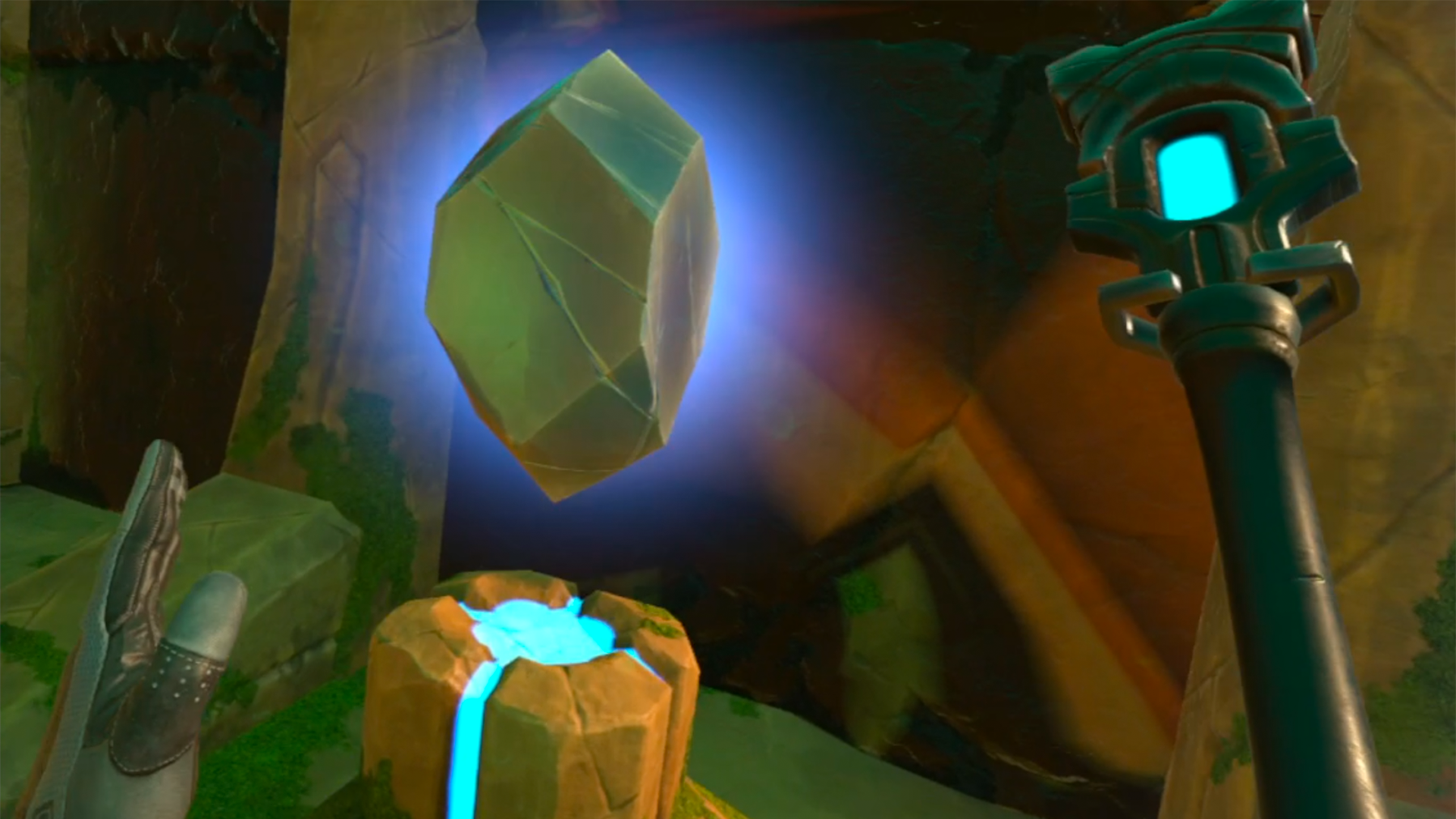}
  \caption{%
    Images from the two VR games that we developed by modifying freely available Unity Learn games. Our VR karting game is on the left, and our VR adventure game is on the right.
  }
  \label{fig:teaser}
}

\abstract{
Research involving virtual reality (VR) has dramatically increased since the introduction of consumer VR systems. In turn, research on VR games has gained popularity within several fields. However, most VR games are closed source, which limits research opportunities. Some VR games are open source, but most of them are either very basic or too complex to be easily used in research. In this paper, we present two source-available VR games developed from freely available Unity Learn games: a kart racing game and a 3D adventure game. Our hope is that other researchers find them easy to use for VR studies, as Unity Technologies developed the games for beginners and has provided tutorials on using them.
} 


\usepackage[normalem]{ulem}

\newcommand{\cut}[1]{}
\newcommand{\cutcite}[1]{}

\begin{document}


\firstsection{Introduction}

\maketitle

Virtual reality (VR) has become broadly popular and accessible since the introduction of consumer VR systems like the HTC Vive and Oculus Rift \cite{Rizzo2017Primetime}. As a result, VR has been applied to a wide range of fields, including industrial simulations, education, public health, social interaction, and entertainment \cite{Hamad2022VRChangedOurLives}. Likewise, there has been a dramatic surge in the number of peer-reviewed publications with ``virtual reality'' as a keyword \cite{Vasser2020VRinPsychologicalResearch}.

Research on VR games, both serious and for entertainment, has also dramatically increased in recent years. Serious VR games have been investigated for biology, chemistry, engineering, nursing, and more \cite{Radianti2020VRforEducation}. Similarly, research on VR games for entertainment has investigated how many games are released each month, their quality, headset support, and complaints of cybersickness \cite{Epp2021PopularVRGames}.

A key issue that limits research on VR games is that most consumer games are closed source, which prohibits them from being modified to facilitate research, such as automatic data collection. \I{CLOVR}, a tool for collecting and logging data from any OpenVR-based application like \I{Beat Saber} and \I{Half-Life: Alyx}, was recently introduced \cite{SegarraMartinez2024CLOVR}. However, most researchers end up developing their own VR games, which are often rudimentary due to constraints.

There are some preexisting VR games that are open source, which allows for researchers to modify them as needed. Ghrairi et al. \cite{Ghrairi2018VRGitHub} found and analyzed 320 open source VR projects on GitHub, of which 37 were games. However, many of the game projects were produced by amateur developers, student VR clubs, or during hackathons. As a result, many of the games are rudimentary in nature and not sufficiently documented for research purposes. On the other hand, some consumer VR games like \I{First Hand} have been open sourced\footnote{\url{https://youtu.be/bUKY6H7_MHw}} but are complex to work with.

In this paper, we present two source-available VR games developed from freely available (non-VR) Unity Learn games: 1) \I{Karting Microgame}---a kart racing game, and 2) \I{The Explorer: 3D Game Kit}---a 3D adventure game. Both of these games are provided by Unity Technologies under its Unity Companion License, which permits derivative works and distribution, subject to terms and conditions\footnote{\url{https://unity.com/legal/licenses/unity-companion-license}}. Most importantly, Unity Learn provides beginner tutorials on how to work with these games, which should facilitate researchers in modifying them for studies. 

\section{Karting Microgame}

The \I{Karting Microgame} is a 3D kart racing game. The Unity Asset Store package provides modular track prefabs to enable developers to quickly create karting tracks\footnote{\url{https://assetstore.unity.com/packages/templates/unity-learn-karting-microgame-urp-150956}}. Unity Learn also provides a 30-minute beginner tutorial on how to customize the game\footnote{\url{https://learn.unity.com/project/karting-template}}.

In order to make a VR version of the \I{Karting Microgame} that is useful for research purposes, we used the SteamVR Unity Plugin, which is openly provided by the Valve Corporation under a BSD 3-Clause ``New'' or ``Revised'' License\footnote{\url{https://github.com/ValveSoftware/steamvr_unity_plugin}}. Due to the \I{Karting Microgame} being built with Unity's Universal Render Pipeline (URP), we first had to use Unity's Render Pipeline Converter feature to convert the SteamVR assets, which were made for the Built-in Render Pipeline, to be compatible with URP. Then, for each game scene, we replaced the main camera and Cinemachine camera controller with the ``Player'' prefab from SteamVR's Interaction System, which was attached to the kart's body. This afforded the ability to view the game from the perspective of driving the cart using a VR headset, such as the Meta Quest 3.

The kart was originally controlled using a conventional WASD scheme and keys. To afford greater interaction fidelity \cite{McMahan2016InteractionFidelity}, we replaced the functionality of the left and right keys (i.e., the A and D keys, respectively) by extending SteamVR's ``CircularDrive'' script to a steering wheel game object, which allows the user to steer the kart by using their VR handheld controllers. We replaced the functionalities of the W and S keys with SteamVR actions for ``Accelerate'' and ``Brake'', which can be dynamically mapped to VR controller buttons, based on the VR system used.

User interface (UI) components were converted from screen space to world space for viewing them within the VR headset. Colliders were added to individual buttons to afford interactions with the UI elements. The ``SteamVR\_LaserPointer'' script was added to the controllers to enable pointing at and selecting the UI elements. We also added a virtual keyboard to the ``IntroMenu'' scene to enable the player to enter their name. 

By default, the \I{Karting Microgame} only involves racing a single lap on its ``Oval Track'' while attempting to complete the lap within 60 seconds and earning additional time by passing track checkpoints. To provide a more in-depth gaming experience, we replicated the main racing scene three times to also incorporate the three other tracks available within the project: ``Country Track'', ``Winding Track'', and ``Mountain Track''. For each scene, the checkpoints were re-positioned to align to the new track. A submenu was added to the ``IntroMenu'' scene to afford selection of these new track scenes. Additionally, Unity PlayerPrefs were used to keep track of the best three laptimes for each track, and the existing leaderboard functionality was updated to display times for the chosen track. 

By modifying the \I{Karting Microgame}, we have created a source-available VR racing game that researchers can use and extend for a multitude of purposes. Some potential research uses include investigating skill progression, cybersickness, or the effects of adding interaction cues. Our VR kart racing game is available on GitHub\footnote{\url{https://github.com/VTXraiLab/VR_Karting_Microgame}}.

\section{The Explorer: 3D Game}

\I{The Explorer: 3D Game Kit} is a two-level, 3D adventure game. The Unity Asset Store package provides a collection of prefabs, tools, and systems to enable developers to quickly create their own adventure game without writing any code\footnote{\url{https://assetstore.unity.com/packages/templates/tutorials/unity-learn-3d-game-kit-115747}}. Unity Learn also provides a 5-hour beginner tutorial on how to use the kit\footnote{\url{https://learn.unity.com/project/3d-game-kit}}.

We again used the SteamVR Unity Plugin to make a VR version of the game that would be useful for research purposes. We replaced the original Ellen player character with SteamVR's ``Player'' prefab. This afforded the ability to view and walk around the game using a VR headset. 

The game originally used the conventional WASD keys to control the horizontal movements of the player. We replaced the functionalities of these keys by developing a ``Locomotion'' script that affords head-directed steering, in which movements are applied to the player object by multiplying the joystick vector by the camera's forward vector. We also used SteamVR's ``SnapTurn'' script to allow the user to virtually turn within the game scene in 45° increments. Finally, we replaced the game's original jump functionality, which was conventionally activated using the space bar, with a SteamVR action for ``Jump'', mapped to one of the VR controller buttons. Similar to our locomotion method, our jump method applies an upward movement to the player object and uses its collider to determine if the player can jump. Gravity in the game brings the player object back to the ground after a jump.

In addition to updating the controls for player movements, we also had to apply the game's original ``Player'' layer to the new SteamVR player object. Enemies (i.e., alien monsters) specifically look for objects on the ``Player'' layer to determine what to attack. Furthermore, the game contains several triggers in the form of pressure plates and switches that only react to ``Player'' layer objects.

We also had to majorly update the attack methods within the game. Originally, animations were used for picking up a staff weapon and swinging it to attack enemies or break boxes. First, we added a SteamVR ``Interactable'' script to the staff and implemented a SteamVR skeleton pose on its handle for picking up the weapon with a virtual hand. We then implemented a box collider around the head of the weapon and created a script for sending damage messages to objects with a ``Target'' tag when collisions of a prerequisite velocity occurred with the staff head.


By modifying \I{The Explorer: 3D Game Kit}, we have created a source-available VR adventure game that researchers can use for several purposes. Some potential research uses include investigating wayfinding behaviors, the effects of different virtual locomotion techniques, and game engagement. Our VR adventure game is available on GitHub\footnote{\url{https://github.com/VTXraiLab/VR_3D_Game_Kit}}.

\section{Conclusion}

In this paper, we have presented two source-available VR games developed from freely available Unity Learn games: a VR kart racing game and a VR adventure game. As a result, these games are more-developed than some amateur VR games that have been openly available. On the other hand, these games are also intended for Unity beginners, which makes them easier to modify than some professional-grade VR games that have been open sourced. Furthermore, tutorials for modifying and extending these games are available from Unity Learn. In conclusion, researchers should find our games useful for implementing VR game-based studies.

\bibliographystyle{abbrv-doi}

\bibliography{template}
\end{document}